\documentstyle[preprint,aps]{revtex}

\def\journal#1, #2, 1#3#4#5, #6{
    {\sl #1~}{\bf #2} (1#3#4#5) #6}
\def\pr{\journal Phys. Rev., }

\def\prl{\journal Phys. Rev. Lett., }

\def\cmp{\journal Comm. Math. Phys., }
\def\np{\journal Nucl. Phys., }

\def\jmp{\journal J. Math. Phys., }
\def\jp{\journal J. Phys., }
\def\beq{\begin{equation}}
\def\eeq{\end{equation}}
\def\ba{\begin{eqnarray}}
\def\ea{\end{eqnarray}}
\def\nn{\nonumber \\}
\def\l{\lambda}
\def\o{\omega}

\newcommand{\cl}{Calogero}
\newcommand{\h}{Hamiltonian}
\newcommand{\col}{collective}
\newcommand{\pv}[1]{{-  \hspace {-4.0mm} #1}}

\begin{document}
\draft
\title{Solitons in the \cl-Sutherland Collective-Field Model}
\author{I. Andri\'c, V. Bardek and L. Jonke\footnote{e-mail address:
andric@thphys.irb.hr \\ \hspace*{3cm} bardek@thphys.irb.hr \\ \hspace*{3cm}
larisa@thphys.irb.hr }}
\address{Department of Theoretical Physics,\\
Rudjer Bo\v skovi\'c Institute, P.O. Box 1016,\\
41001 Zagreb, CROATIA}
\date{\today}
\maketitle
\begin{abstract}
\widetext
In the Bogomol'nyi limit of the \cl-Sutherland \col-field model we find
static-soliton solutions. The solutions of the equations of motion are
moving solitons, having no static limit for $\l>1$. They describe holes and
lumps, depending on the value of the
statistical parametar $\l$.
\narrowtext
\end{abstract}

\newpage

The properties of the \cl-Sutherland model\cite{cal,suth}, which describes
particles in one dimension interacting with a two-body inverse-square
potential, are a subject of increasing interest. The \cl-Sutherland model (CSM)
is integrable in terms of symmetric functions, thus representing a valuable
specific
 model to discuss fractional statistics. Similarities between the CSM and the
two-dimensional
(fractional) quantum Hall system (the Jastrow-type ground-state wave function)
indicate
that the CSM is linked to an anyonic system as the one-dimensional reduction of
the former\cite{iso}.

The CSM also appears when two-dimensional QCD is formulated as a random matrix
model in the large-N limit, restricted to a singlet subspace\cite{a2}.
For a one-dimensional free fermionic Fock space, which is described by a
hermitian matrix model\cite{br}, the \col-field \h\ can be
introduced\cite{ant}.
 The derived effective lagrangian, having the same dispersion relation as the
theory we started from, turns out to be of the \cl\ type with the statistical
parametar $\l=1/2$\cite{ant}.
 This can be seen when the collective-field formulation is written for CSM, as
was done in ref.\cite{a2,a1}. The hole excitations in the spectrum are
represented as a soliton of the \cl-Sutherland \col-field theory.

In this Letter we investigate solitons in the CSM in terms of the \col-field
\h\ description formulated in ref.\cite{a2,a1}. We briefly mention the results
of ref.\cite{a2}, and a similar procedure can be applied to the trigonometric
interaction\cite{suth,a2}.

The \cl\  \h , which describes a system of N non-relativistic particles on a
line interacting via the two-body inverse-square potential, is given by
\beq \label{h1}
H=\frac{1}{2}\sum_{i=1}^Np_i^2 +\frac{\l(\l-1)}{2}\sum_{i\neq
j}^N\frac{1}{(x_i-x_j)^2}.\eeq
The dimensionless coupling constant $\l(\l-1)$ is a positive real number which
specifies the statistics of this model. Because of the singularity of the \h \
for $x_i=x_j$, the wave function ought to have a prefactor which will vanish
for coincident particles. We extract this prefactor in the form
\beq \label{wf1}
\Psi(x_1,x_2,\ldots,x_N)=\prod_{i<j}^{N}(x_i-x_j)^{\l}\Phi(x_1,x_2,\ldots,x_N),
\eeq and obtain the new \h
\beq \label{h2}
H=-\frac{1}{2}\sum_{i=1}^N\frac{d^2}{dx_i^2}-\l\sum_{i\neq
j}^N\frac{1}{x_i-x_j}\frac{d}{dx_i}, \eeq acting on the residual wave function
$\Phi$. The \h\ (\ref{h2}) is now suitable for transformation into a
collective-field representation.
It has been shown that, in the large-N limit, the \h\ can be expressed entirely
in terms of the density of particles $\rho(x)$ and its canonical conjugate
$\pi(x)=-i\frac{\delta}{\delta\rho(x)}$. The Jacobian of the transformation
from $x_i$ into $\rho(x)$ rescales the wave functional
\beq \label{vf}
\Phi(x_1,x_2,\ldots,x_N)=J^{1/2}\Phi(\rho), \eeq
 resulting in the hermitian \col-field \h
\ba \label{h3}
H=&&\frac{1}{2}\int dx\rho(x)(\partial_x\pi)^2+
\frac{1}{8}\int dx\rho(x)\left(\partial_x\frac{\ln
J}{\delta\rho(x)}\right)^2-\nn
-&&\frac{1}{4}\int dx\frac{\delta\o(x)}{\delta\rho(x)}, \ea
with \beq \label{o1}
\o(x)=(\l-1)\partial_x^2\rho(x)+2\l\partial_x\left(\rho(x)\pv{\int}
dy\frac{\rho(y)}{x-y}\right) ,\eeq
 and the Jacobian determined from the hermicity condition
\beq \label{hy}
\partial_x\left(\rho(x)\partial_x\frac{\ln J}{\delta\rho(x)}\right) =\o(x)
.\eeq
The last singular term plays the role of the counter term, and does not give a
contribution in the leading order in N.
To find the ground-state energy of our system, we assume that the corresponding
\col-field configuration is static and has a vanishing momentum $\pi$.
Therefore, the leading part of the \col-field \h\ in the $1/N$ expansion is
given by the effective potential \beq \label{v1}
V_{eff}(\rho)=\frac{1}{8}\int dx\rho(x)\left(\partial_x\frac{\ln
J}{\delta\rho(x)}\right)^2=\frac{1}{8}\int
dx\rho(x)\left[(\l-1)\frac{\partial_x\rho(x)}{\rho(x)}+2\l\pv{\int}
dy\frac{\rho(y)}{x-y}\right]^2 .\eeq
This potential is positive semidefinite and, therefore, its contribution to the
ground-state energy vanishes if there exists a positive solution of the
first-order differential Bogomol'nyi-type equation:
\beq \label{b1}
(\l-1)\frac{\partial_x\rho(x)}{\rho(x)}+2\l\pv{\int}
dy\frac{\rho(y)}{x-y}=0.\eeq The most obvious solution is given by the
constant-density configuration $\rho=\rho_0$ for any value of the statistical
parametar $\l$.
Let us now find an interesting family of ground-state solutions which
represents hole excitations of the \cl\ system. Using the identity for the
principal distribution
\beq \label{xp}
\frac{P}{x-y}\frac{P}{x-z}+\frac{P}{y-x}\frac{P}{y-z}+\frac{P}{z-x}\frac{P}{z-y}=\pi^2\delta(x-y)\delta(x-z), \eeq
and performing partial integration, we can rewrite $V_{eff}$ as
\ba \label{v2}
V_{eff}&&=\frac{1}{8}\int
dx\rho(x)\left[(\l-1)\frac{\partial_x\rho(x)}{\rho(x)}+\frac{2c}{x}+2\l\pv{\int} dy\frac{\rho(y)}{x-y}\right]^2-\nn
&&-\frac{1}{2}c(c-1+\l)\int dx\frac{\rho(x)}{x^2}+\frac{c\l}{2}\left(\int
dx\frac{\rho(x)}{x}\right)^2-\nn
&&-\frac{c\l}{2}\pi^2\rho^2(0). \ea
We are looking for the symmetric minimum $\rho(x)=\rho(-x)$, representing a
hole located at the origin $\rho(0)=0$. For the particular value of the
constant $c$ given by \beq \label{c2} c=1-\l ,\eeq the Bogomol'nyi limit
appears. The contribution of $V_{eff}$ vanishes and the corresponding
configuration satisfies the Bogomol'nyi equation
\beq \label{b2}
(\l-1)\frac{\partial_x\rho(x)}{\rho(x)}+\frac{2(1-\l)}{x}+2\l\pv{\int}
dy\frac{\rho(y)}{x-y}=0. \eeq
We now see that the role of the new singular term in the equation is to
compensate for the singularity produced by $\partial_x\ln\rho(x)$ at the point
where the \col\ field $\rho$ vanishes. Equation (\ref{b2}) can be solved by a
rational ansatz
\beq \label{r3}
\rho(x)=\frac{ax^2}{b^2+x^2}. \eeq
Inserting the ansatz for $\rho(x)$ (\ref{r3})
into our equation (\ref{b2}), using the Hilbert transform
\beq \label{ht}
\pv{\int}\frac{dy}{x-y}\frac{1}{a^2+y^2}=\frac{\pi}{a}\frac{x}{a^2+x^2},\eeq
and after performing some calculation we find the condition
\beq \label{c1}
 ab\pi=\frac{1-\l}{\l} .\eeq An acceptable positive solution exists only for
$\l<1$. For large values of x, the soliton solution (\ref{r3}) approaches to
the constant solution found before. It can be shown that the net particle
number carried by the our soliton $\rho(x)$ is
\beq \label{n1}
\int dx (\rho(x)-\rho_0)=\frac{\l-1}{\l}<0. \eeq
This fact indicates again that our soliton corresponds to the hole excitation.
There is a correspondnig prefactor in front of the wave function which
describes
 the hole. Generally, adding a term inside the bracket (as was done in
(\ref{v2})), means adding a new term to the logarithm of the Jacobian. The
corresponding prefactor, which arises because of adding $(1-\l)/(x-X)$,
 is $\prod_i(x_i-X)^{1-\l}$, describing the hole at place $X$.

So far we have considered soliton solutions originating from the Bogomol'nyi
lower bound on the ground-state energy. Let us now turn our attention to the
possible solutions which cannot be reached by the Bogomol'nyi saturation. To
find
such a solution, it is necessary to minimize the \col\ \h\ with respect to
$\rho$ i $\pi$, i.e. to find the corresponding dynamical equations of motion.
In this case, some interesting investigations have already been done in the
recent
literature\cite{ant}. We shall rederive the solution studied by
the author of ref.\cite{ant}, to obtain a general description, valid for any
value of the statistical parametar
$\l$.

The equations to be solved are
\begin{mathletters}
\beq \label{jg1}
\dot{\rho} =-\partial_x(\rho(x)\partial_x\pi), \eeq
\beq \label{jg2}
\dot{\pi} =-\frac{1}{2}(\partial_x\pi)^2-\frac{\delta V_{eff}}{\delta\rho},
\eeq
\end{mathletters}
where $V_{eff}$ is given by (\ref{v1}). Since we are looking for
constant-profile solutions, propagating at speed $v$, depending only on $\xi
=x-vt$, we obtain
\begin{mathletters}
\beq \label{jg3}
\frac{d\pi}{d\xi}=v\left( 1-\frac{\rho_0}{\rho}\right) , \eeq
\beq \label{jg4}
v^2\left( 1-\frac{\rho_0}{\rho}\right)=\frac{v^2}{2}\left(
1-\frac{\rho_0}{\rho}\right)^2+\frac{\delta V_{eff}}{\delta\rho},
\eeq\end{mathletters}
where $\rho_0$ is a constant solution defined by
\beq \label{m}
\frac{\pi^2\l^2}{2}\rho_0^2=\mu . \eeq
Here, we have implemented the Lagrange multiplier $\mu$, which defines the
energy scale of the problem.
Some algebra finally leads to the equation for a moving soliton-solution:
\ba \label{ms}
\frac{v^2}{2}\left(\frac{\rho_0^2}{\rho^2}-1\right)
+\frac{\pi^2\l^2}{2}(\rho^2-\rho_0^2)-\frac{(\l-1)^2}{4}\partial_{\xi}\left(\frac{\partial_{\xi}\rho}{\rho}\right) - \nn  -\frac{(\l-1)^2}{8}\left(\frac{\partial_{\xi}\rho}{\rho}\right) ^2 -\l(\l-1)\partial_{\xi}\pv{\int} d\eta\frac{\rho(\eta)}{\xi -\eta} =0 . \ea
Plugging the rational ansatz for $\rho(\xi)$ into (\ref{ms}):
\beq \label{r4}
\rho(\xi)=\frac{\rho_0\xi^2+a^2}{\xi ^2+b^2} , \eeq
we find the following condition on the parametars $a$ and $b$:
\begin{mathletters}
\beq \label{c3}
b=\frac{\l(\l-1)\pi\rho_0}{v^2-\rho_0^2\pi^2\l^2} , \eeq
\beq \label{c4}
a^2=b^2\rho_0+\frac{\l-1}{\l\pi}b=\frac{(\l-1)^2
v^2\rho_0}{[v^2-\rho_0^2\pi^2\l^2]^2}. \eeq
\end{mathletters}
Notice that the above formulae are valid only for $b>0$. Writing the solution
$\rho(\xi)$ in the form
\beq \label{r5}
\rho(\xi)=\rho_0+\frac{a^2-\rho_0b^2}{\xi^2+b^2}, \eeq
it can be easily seen that depending on the sign of the numerator
$a^2-\rho_0b^2$, two basically different soliton profiles emerge. For
$a^2-\rho_0b^2<0,\;(\l<1),$ we have a hole excitation propagating at speed
$|v|<\pi\rho_0\l$.
 For $a^2-\rho_0b^2>0,\;(\l>1)$, we have a lump solution propagating at speed
$|v|>\pi\rho_0\l$. The static $(v=0)$ soliton exists only for $\l<1$, and
corresponds to the solution found before (\ref{r3}). The net particle number
carried by the soliton $\rho(\xi)$ is defined by
\beq \label{n2}
\Delta N=\int d\xi(\rho(\xi)-\rho_0)=(a^2-\rho_0b^2)\int
\frac{d\xi}{\xi^2+b^2}=\frac{\l-1}{\l} . \eeq

The energy of the moving soliton is defined with respect to the stationary
background given by $\rho_0$
\beq \label{en}
E=H(\rho(\xi))-H(\rho_0) . \eeq
Using the equation of motion (\ref{ms}), we obtain
\beq \label{ee}
E=\frac{\l-1}{\l}\left(\frac{v^2}{2}-\mu\right) . \eeq  Let us now find the
momentum of the moving soliton. The \col-field theory gives
\beq \label{pp}
P=\sum_ip_i=\int dx\partial_x\rho(x)\pi(x) . \eeq
After partial integration, the momentum (\ref{pp}) takes the form
\beq \label{pe}
P=-\int dx\rho(x)\partial_x\pi+\rho_0(\pi(\infty)-\pi(-\infty)).
\eeq  Inserting the expression (\ref{jg3}) for $\partial_x\pi$, we easily
obtain
\beq \label{p}
P=\frac{\l-1}{\l}(\sqrt{2\mu} -v) . \eeq
{}From the relations (\ref{ee}) and (\ref{p}), we have the following dispersion
law for the moving soliton:

\beq \label{dis}
E(P)=\frac{\l}{\l-1}\frac{P^2}{2}+\sqrt{2\mu}|P| . \eeq

We conclude by noting that our moving-soliton solution exists only for $\l$
different from zero or one, i.e. for generic intermediate statistics. The
effect of
including quantum fluctuations around the moving solitons and its implications
on the dispersion law will be considered elsewhere.

{\it Note added}. After this work was completed, we became aware of
ref.\cite{pol}. The moving-soliton solution found there corresponds to our
solution (\ref{r5}) only for $\l>1$.
\acknowledgments

One of us (I.A.) is grateful to Antal Jevicki for useful discussion.
This work was supported by the Scientific Fund of the Republic of Croatia.

\end{document}